\RequirePackage{amsmath}

\documentclass[a4paper, 10pt, conference]{IEEEtran}

\usepackage[utf8]{inputenc}
\usepackage{graphicx}
\usepackage{amssymb}
\usepackage{setspace}
\usepackage{rotating}
\usepackage{algorithm}
\usepackage{algorithmicx}
\usepackage[noend]{algpseudocode}
\usepackage[export]{adjustbox}
\usepackage[biblabel]{cite}
\usepackage{graphicx}
\usepackage{float}
\usepackage{xcolor}

\begin{document}
               
\title{Assessment of the Relative Importance of different hyper-parameters of LSTM for an IDS}

\author{\IEEEauthorblockN{Mohit Sewak}
\IEEEauthorblockA{\textit{
Security \& Compliance Research}\\
\textit{Microsoft, India}\\
mohit.sewak@microsoft.com}
\and
\IEEEauthorblockN{Sanjay K. Sahay}
\IEEEauthorblockA{\textit{Department of CS \& IS, Goa Campus} \\
\textit{BITS Pilani, Goa, India}\\
ssahay@goa.bits-pilani.ac.in}
\and
\IEEEauthorblockN{Hemant Rathore}
\IEEEauthorblockA{\textit{Department of CS \& IS, Goa Campus} \\
\textit{BITS Pilani, Goa, India}\\
hemantr@goa.bits-pilani.ac.in}
}

\maketitle  
\begin{abstract}
Recurrent deep learning language models like the LSTM are often used to provide advanced cyber-defense for high-value assets. The underlying assumption for using LSTM networks for malware-detection is that the op-code sequence of a malware could be treated as a (spoken) language representation.
There are differences between any spoken-language (sequence of words/sentences) and the machine-language (sequence of op-codes).
In this paper we demonstrate that due to these inherent differences, an LSTM model with its default configuration as tuned for a spoken-language, may not work well to detect malware (using its op-code sequence) unless the network's essential hyper-parameters are tuned appropriately. In the process, we also determine the relative importance of all the different hyper-parameters of an LSTM network as applied to malware detection using their op-code sequence representations.
We experimented with different configurations of LSTM networks, and altered hyper-parameters like the embedding-size, number of hidden-layers, number of LSTM-units in a hidden layers, pruning/padding-length of the input-vector, activation-function, and batch-size. 
We discovered that owing to the enhanced complexity of the malware/machine-language, the performance of an LSTM network configured for an Intrusion Detection System, is very sensitive towards the number-of-hidden-layers, input sequence-length and the choice of the activation-function. Also, for (spoken) language-modeling, the recurrent architectures by-far outperforms their non-recurrent counterparts.
Therefore, we also assess how sequential DL architectures like the LSTM compares against their non-sequential counterparts like the MLP-DNN for the purpose of malware-detection.
\end{abstract}

\section{Introduction} \label{sec:introduction}
Various static and dynamic analysis has been conducted using different features for malware detection. Since static analysis is more practical for production usage, a lot of research is focused in this direction. While the first generation malware could be detected by simple signature matching, more advanced polymorphic and metamorphic could be better detected only by using machine learning (ML), and deep learning (DL) approaches\cite{sahay}.
Many features have been used for static analysis of malware. Of these, methods based on op-code analysis, especially the op-code-frequency feature vector analysis, have achieved good success. The study by Santos et. al. \cite{1-Santos-2013} claims to have obtained an accuracy of $95.9\%$ using op-code frequency as features on a standardized malware dataset. Later, researchers also applied file-size based segmentation over the op-code frequency vector and improved the average accuracy over different ML models slightly over $96\%$, and claimed that random forest based Intrusion Detection System (IDS) provided the best individual accuracy close to $98\%$ with a False Positive Rate (FPR) of $1.07\%$ \cite{2-Ashu-RF} on the same standardized malware data \cite{3-Malicia}.
Later on Sewak et. al. \cite{sewak-dnn} improved both the accuracy and False Positive Rate (FPR) of malware detection on the same dataset using similar op-code frequency feature vectors but by using DL methods like the Multi Layer Perceptron (MLP) based Deep Neural Networks (DNN). Further, Sewak et. al. also compared the DL/DNN based IDS with corresponding ML based IDS \cite{sewak-comparison}. Sewak et. al. did use unsupervised DL techniques like Auto Encoders (AE) for feature extraction and thereby achieved improved accuracy of $99.21\%$ with even a better FPR of $0.19\%$. So far the work by Sewak et. al. has produced the best performance on the standardized Malicia data-set.
In this paper we extend the work of Sewak et. al. and analyze the performance while using op-code sequence vector instead of frequency vector could help achieve better results. Since a DNN based network cannot efficiently analyze sequence data, therefore we use Recurrent Neural Network (RNN) architecture based Long Short Term Memory (LSTM) networks. We use different combination of LSTM architectural parameters to ensure that the work is not biased towards a particular network architecture. 
The main contributions of this paper are as follows:

\begin{itemize}
    \item Analysing the relative importance of each of the below architectural-parameters for an LSTM based malware-detection system:
    \begin{itemize}
        \item Sequence Length
        \item Embedding Size
        \item Number of LSTM Layers
        \item Number of units in LSTM hidden-layers
        \item Activation function
        \item Dropout ratio
        \item Batch Size
    \end{itemize}
    \item Determining that the number-of-LSTM-layers and activation-function are the most important criteria, followed by the length of input-sequence and embedding-size. This indicates that the malware-detection using op-code sequence is a complex problem. Also, the op-code features important for detection could be further apart in the call sequence and may not have a trivial connection.
    \item Determine how does the two DL based IDS, i.e. an LSTM based IDS and a MLP-DNN based IDS compares for the purpose of malware detection. 
\end{itemize}

To the best of our knowledge such exhaustive comparison on LSTM's hyper-parameters' relative importance for malware detection is not available in any known literature, nor is a direct comparison between any sequential and non-sequential DL architecture for IDS. The flow of the remaining paper is as follows. In section \ref{sec:realted-work} we cover some related work on malware detection and use of LSTM/ RNN (in sub-section \ref{sec:relatedwork-lstm}). Then in section \ref{sec:malware-data-used} we describe the malware data used. The pre-processing and cleaning of this data is covered in section \ref{sec:data-preprocessing}, followed by LSTM modeling in section \ref{sec:modeling}. We then present our results in section \ref{sec:results} and discuss their significance in section \ref{sec:discussion}; and finally conclude the paper in section \ref{sec:conclusion}.

\section{Related work} \label{sec:realted-work}
A lot of recent efforts has been focused on, and much success achieved with different DL techniques for malware detection. Some DL models used for making IDS are the Deep Belief Networks \cite{20-DeepSign-2015}, Deep Neural Networks \cite{21-Saxe-2015}\cite{hemant-bda}. RNN and its variants like Echo State Networks (ESN)\cite{19-Pascanu-2015} LSTM \cite{opcode-lstm-2layer} and Gated Recurrent Unit (GRU) \cite{microsoft-lstm} and combination of recurrent and convolutional neural networks \cite{22-Kolosnjaji-2016} for supervised learning and classification. RNN based Auto-Encoders (RNN-AE) \cite{18-Xin-2016} have also been used as approaches for feature generation for downstream supervised learning mechanism.
Many of the work on RNN/ LSTM based models above \cite{microsoft-lstm, opcode-lstm-2layer} use a language representation of the malware as reflected in the extracted op-code sequences of the candidate file. \newline
LSTMs have been used in the field of Malware detection on various types of features. Salient amongst them being the use of API call sequences \cite{LSTM-api-call}, System call sequences \cite{LSTM-system-call} file headers \cite{LSTM-PE-header} and opcodes \cite{opcode-lstm-2layer}\cite{hemant-cluster}. Such implementations have been used for platforms ranging from Windows PEs \cite{LSTM-PE-header}, to Android \cite{LSTM-system-call}, to IoT devices \cite{LSTM-iot}. The application of such techniques is targeted towards purposes ranging from detection of malicious files to detection of a specific malware malware \cite{LSTM-api-call} to identification of obfuscation in malware \cite{LSTM-obfuscation}.\newline
In some work that use LSTM on op-code sequences \cite{LSTM-HDN} it has been acknowledged that the since op-code sequences are much longer than language representations, so they may not work ideally as-is for such features. In some other works it was found that LSTM with more number of layers work better for malware detection \cite{opcode-lstm-2layer, LSTM-iot}. But no prior work exists that assess the relative importance of different configurable architectural and hyper-parametric settings for an IDS. Also, no prior work exist that compares an MLP DNN based IDS directly to an LSTM based one on a standardized dataset.

\subsection{About LSTM Network and LSTM Unit} \label{sec:relatedwork-lstm}
 An RNN network has directed temporal linkages to memory states of cells in its previous sequence-step. Because of this feature they can improve output representation with information present in temporally lagged sequences. This property makes the RNN networks (and its variants, like the LSTM) useful for applications like speech-recognition, language-modeling etc. This is because correct output representations in such applications also depends on information in previous sequence-steps. 
 Unfortunately, basic RNN suffers from the \textit{Exploding} and \textit{Vanishing Gradient} problems like the classical Artificial Neural Network (ANN) suffered before the invent of DL and activation functions like \textit{Rectified Linear Unit} (ReLU). Because of these problems the RNNs in its native form could not be effectively made deep (having multiple hidden layers) enough and hence could not discover complex patterns in the sequential data.
 LSTM \cite{lstm-original-paper}, a variant of RNN, solves this problem by using multiple memory gates. An LSTM unit consists of a cell, an input gate, an output gate, and a remember gate. Later a forget-gate \cite{lstm-forget-gate} was added to the LSTM architecture so that it could reset its state completely or partially. The forward-pass of an LSTM with the forget gate could be expressed mathematically as below \cite{lstm-forget-forward-pass}.
 \begin{equation}
     f_t = \sigma_g(W_fx_t + U_fh_{t-1} + b_f)
    \label{eqn:lstm-forget}
 \end{equation}
  \begin{equation}
     i_t = \sigma_g(W_ix_t + U_fh_{t-1} + b_i)
     \label{eqn:lstm-input}
 \end{equation}
  \begin{equation}
     o_t = \sigma_g(W_ox_t + U_oh_{t-1} + b_o)
     \label{eqn:lstm-output}
 \end{equation}
  \begin{equation}
     c_t = f_t \circ c_{t-1} + i_t \circ \sigma_c(W_cx_t + U_ch_{t-1} + b_c)
     \label{eqn:lstm-cell}
 \end{equation}
  \begin{equation}
     h_t = o_t \circ \sigma_h(c_t)
     \label{eqn:lstm-hidden}
 \end{equation}
 Where $x_t \in \mathbb{R}^d$ is the input vector to the LSTM unit. In equation \ref{eqn:lstm-forget}, $f_t \in \mathbb{R}^h$ is the \textit{forget} gate's \textit{activation-vector} at step `t'. In equation \ref{eqn:lstm-input}, $i_t \in \mathbb{R}^h$ is the \textit{input} gate's \textit{activation-vector} at step `t'.  In equation \ref{eqn:lstm-output}, $o_t \in \mathbb{R}^h$ is the `output' gate's \textit{activation-vector} at step `t'.  In equation \ref{eqn:lstm-cell}, $c_t \in \mathbb{R}^h$ is the \textit{cell's state-vector} at step 't'. In equation \ref{eqn:lstm-hidden}, $h_t \in \mathbb{R}^h$ is the \textit{output} or \textit{hidden state-vector} at step `t'.
 The $W\in\mathbb{R}^{h\times d}$, and $U\in \mathbb {R} ^{h\times h}$ represents the weight-matrices and $ b\in \mathbb {R} ^{h}$ represents the bias-vector parameters which need to be learned during training. `d' and `h' represents the number of input features and number of hidden units.
 In the equation \ref{eqn:lstm-cell} and \ref{eqn:lstm-hidden} the symbol `$\circ$' represents the \textit{Hadamard-Product} \cite{lstm-hadamard-product} which is a form of an element-wise product. All the gates (forget, input, output) uses the \textit{Sigmoid Activation} \cite{sewak-dl-overview} represented by the symbol `$\sigma_g$', whereas the cell state-vector use the \textit{hyperbolic-tangent} (tanh) activation-function \cite{sewak-dl-overview} represented by symbol `$\sigma_c$'. The hidden/output state-vector use another form of hyperbolic-tangent function \cite{lstm-peephole-activation} represented by the symbol `$\sigma_h$', given as $\sigma_h(x)=x$.

\section{Malware Data-set} \label{sec:malware-data-used}
For our experiments, the malware data was collected from the Malicia \cite{3-Malicia} project. The project contains $11,368$ windows Portable Executable (PE) malicious files. From all the PE, their respective assembly (*.asm) files were created using Unix’s \textit{objdump} utility, which succeeded in disassembling $11,084$ of the files. 
Another sample of $2,819$ benign executable files was collected from different windows systems using the \textit{Cygwin} utility. Each supposedly non-malicious file in the sample was verified to be non-malicious using the \textit{virustotal.com} file scanner services. Op-code sequences feature vectors were extracted by disassembling all these PEs.

\subsection{Data Pre-Processing and Cleaning} \label{sec:data-preprocessing}
\begin{figure*}[htbp]
    \centering
    \includegraphics[width=\textwidth]{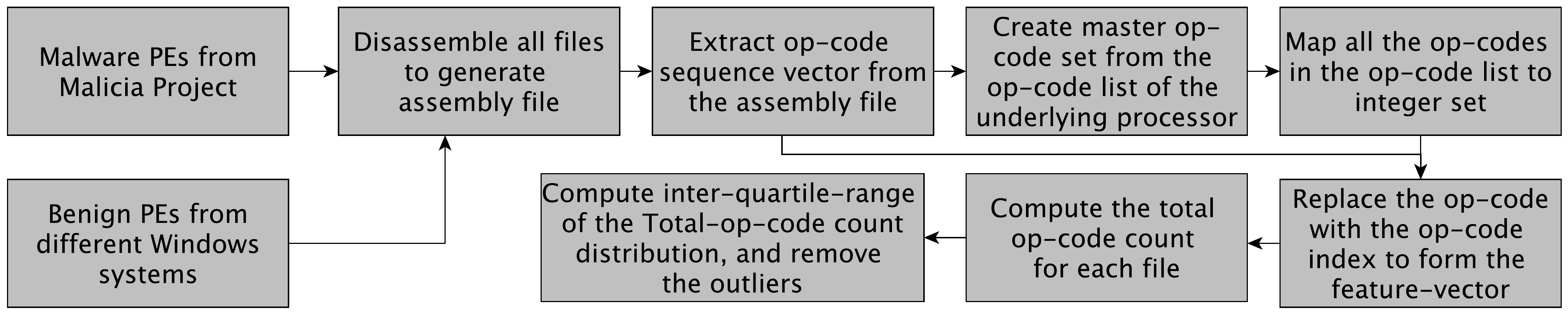}
    \caption{Process flow for extracting feature vectors}
    \label{fig:data-processing-flow}
\end{figure*}
\begin{figure*}[htbp]
    \centering
    \includegraphics[width=\textwidth, height=2.4in]{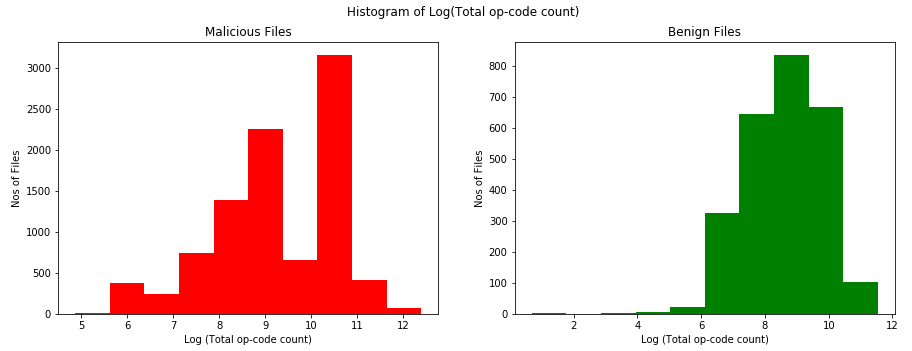}
    \caption{Histogram of $Log_{10}(opcode-sequence-length)$ in Malicious and Benign files }
    \label{fig:data-malicious-opcode-number-histogram}
\end{figure*}

Data pre-processing is carried out as indicated in the data processing flow as shown in figure \ref{fig:data-processing-flow}. We map all the $1612$ unique op-codes in the instruction-set of the underlying processing architecture to a contiguous sparse index vector. This (sparse) integer index vector was in-turn later mapped into a lower dimension dense embedding tensor as DL works best with such data-structures and sparse categorical data representations (like the op-code name) is not suitable for training such models.
Next, the histogram of the total op-code count was plotted to observe the distribution of op-code sequence-length across files in both the classes (malicious and benign). This is as shown in figure \ref{fig:data-malicious-opcode-number-histogram}. It was observed that the op-code sequence-vector had huge variance in sequence-length. For application in LSTM we need a consistent length cut-off, so this variance needs to be addressed. The malicious and non-malicious PEs op-code sequence vectors were recursively extracted and the total count of all op-code computed for each candidate file.
Since the distribution of op-code count (figure \ref{fig:data-malicious-opcode-number-histogram}) is highly right-skewed therefore outliers need to be identified and removed to have a consistent data for training the models. Since we have fewer non-malicious data than malicious, we wanted to be more conservative in pruning the non-malicious data.
To identify the outliers quantitatively, different percentiles of the total count of op-codes in non-malicious and malicious files were observed. Table \ref{tbl:quantile-op-code-series} shows the different percentile markers for non-malicious and malicious data. Based on the observations from this histogram, and percentile mapping, we remove the files with null data and less than $1$ percentile of non-nulls. 

Additionally, more scientific methodology of inter-quantile range was used. Since the sequence-length distribution of malicious files is exponentially right-skewed than non-malicious files we modify the standard inter-quartile method as shown in equation \ref{eqn:opcode-max-threshold}.
\begin{equation}
     MaxOpCodeThreshold = \mathbb{E}(O) + 1.0*(\mathbb{Q}3(O) - \mathbb{Q}1(O))
    \label{eqn:opcode-max-threshold}
\end{equation}
In \ref{eqn:opcode-max-threshold}, the symbol $O$ represents the op-code count vector for different files, and $\mathbb{Q}$ representations represents quartiles of this series. This process was repeated until all the outliers were removed. We selected a (nearest hundred) clipping mark of 100 minimum total op-codes for retaining the malicious files in the experiment. We were finally left with $1,139$ non-malicious files, and $8,116$ malicious files for training.
The maximum op-code count in a file even after outlier reduction as discussed in section \ref{sec:data-preprocessing} were $~44,000$. This is still too high to be processed recursively in a recurrent DL network. Also, the total op-code counts, and hence the op-code sequence length, across files was quite heterogeneous. Such heterogeneity in op-code sequence length is not conducive for use in a recurrent DL network. To resolve this issue we conducted the \textit{trim-pad} operation to the op-code sequence vector in each file. We chose a constant length of op-code sequence for a given run/ experiment of the model training. This length ($L_{\textit{sequence}}$) is kept static for a given model configuration run. We use the value of $\mathbb{E}(O)$ and $\mathbb{Q}3(0)$ for setting the $L_{\textit{sequence}}$.
\begin{table}[htbp!]
    \begin{center}
     \begin{tabular}{||c c c || c c c||} 
     \hline
     \%ile & NonMal & Mal & \%ile & NonMal & Mal\\ [0.5ex] 
     \hline\hline
        01    &  186  &      0 & 75    & 14385 & 34498\\
        05    &  725  &      0 & 90    & 24464 & 45049\\
        10    & 1069  &   503  & 95    & 31261 & 51709\\
        25    & 2336  &  3454  & 99    & 50258 & 84237 \\
        50    & 7327  &  9884  & 100   & 63943 & 286305\\
     \hline
     \hline
    \end{tabular}
    \end{center}
    \caption{Percentile from distribution length of op-code sequence vector across malicious and non-malicious files}
    \label{tbl:quantile-op-code-series}
\end{table}

\section{LSTM configurations used and modeling approach} \label{sec:modeling}
\begin{figure*}[htb]
    \centering
    \includegraphics[width=\textwidth, height=2.5in]{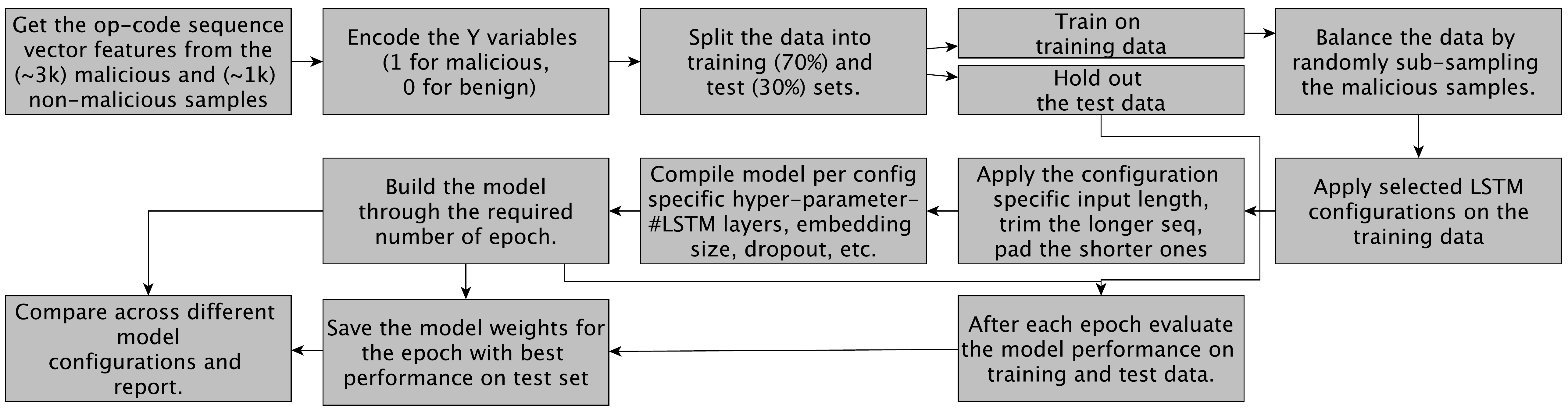}
    \caption{Process Flow for modeling}
    \label{fig:lstm-modeling-flow}
\end{figure*}

Once we pre-process the data, removed outliers from the dataset based on their op-code sequence-length, and homogenized the input sequence-length for the remaining data (by trimming/padding the op-code sequences) as covered in section \ref{sec:data-preprocessing}, we process the data-set for modeling. The process flow for modeling is as shown in figure \ref{fig:lstm-modeling-flow}. 
We assign the labels to the dataset by encoding malware as `1' and non-malicious file-features as `0'. We then split the data-set into training and test/validation datasets. 
 The data-set was then partitioned into training and test/validation set.
 The final dataset after the pre-processing and outlier removal contains $~11,000$ malicious and $~2,800$ benign samples. This disparity in samples across the classes leads to class imbalance problem. The class imbalance problem was resolved by randomly under-sampling the malicious data.
 Next, we proceed to applying the different LSTM configurations over the training set and evaluating the model thus trained on the test set to generate and compare the results. The setup of the LSTM network architecture common to all the configuration is as described next.
The LSTM network was preceded by an embedding layer in all these experiments for the reasons explained earlier. After the embedding layer, one or more LSTM layers were used. The LSTM layers finally concluded in a \textit{Dense} fully-connected layer which finally terminated into with a single cell output layer, representing the probability of class 1 (malicious). \textit{Sigmoid} activation was used for the output layer and \textit{Binary Cross Entropy} loss computed between the actual target and its prediction probabilities. \textit{Adam} optimizer was used to optimize the losses. The LSTM layers could one of the two types of activation namely the \textit{sigmoid} or the \textit{tanh}. Besides the input and output layer activation, the network also has an \textit{inner activation}, also called the \textit{recurrent activation}. The \textit{hard-sigmoid} function was used as \textit{inner/recurrent} activation in all the configurations. The \textit{drop-out} mechanism was used in each LSTM layer to reduce overfitting while training. In the \textit{drop-out} mechanism a proportion of neurons during training are dropped out (made inactive) to avoid over-fitting. The Drop-Out ratio for the layers were also altered and chosen from a set of \{0.3,0.5\}. A drop-out ratio of 0.0 indicates no drop-out.
Following parameters were altered for the training of different LSTM models and corresponding embedding layers as shown in table \ref{tbl:LSTM-configurations-used}.

\section{Results} \label{sec:results}
\begin{figure}[htb]
    \centering
    \includegraphics[width=\linewidth]{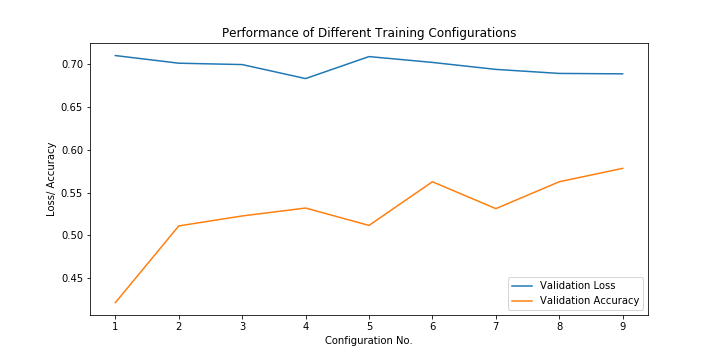}
    \caption{Performance of different training configurations}
    \label{fig:validation-performance}
\end{figure}

The LSTM models with the configurations as shown in table \ref{tbl:LSTM-configurations-used} were trained across 5 to 10 epochs. The training was terminated as soon as convergence was reached and the model weights corresponding to the best (lowest loss) epoch extracted for validation. The validation of the model was done on the held out test-data and the results shown in the later half of the table \ref{tbl:LSTM-configurations-used}. Figure \ref{fig:validation-performance} represents the validation loss and validation accuracy.

\begin{table*}[htbp!]
    \begin{center}
     \begin{tabular}{||c c c c c c c c || c c||}
     \hline
     \multicolumn{8}{c}{Configuration Specification} & \multicolumn{2}{c}{Validation Results}\\
     \hline\hline
     SN & SqLn & EmSz & Lyrs & OutDim & ActFun & DropOut & BchSz & Loss & Acc\%\\ 
     \hline\hline
        1 & Q(0.75) & 128 & 2 & 256 & sigmoid & 0.5 & 128  & 0.7094 & 42.18\\
        2 & Q(0.75) & 128 & 1 & 256 & sigmoid & 0.5 & 128  & 0.7005 & 51.11\\
        3 & Q(0.75) & 256 & 1 & 256 & sigmoid & 0.5 & 128  & 0.6989 & 52.27\\
        4 & Q(0.50) & 128 & 1 & 256 & sigmoid & 0.5 & 128  & 0.6826 & 53.19\\
        5 & Q(0.50) & 256 & 1 & 256 & sigmoid & 0.5 & 128  & 0.7082 & 51.17\\
        6 & Q(1.0) & 64 & 2 & Q(1.0) & tanh & 0.3 & 32     & 0.7014 & 56.25\\
        7 & Q(0.75) & 256 & 2 & Q(0.75) & tanh & 0.3 & 32  & 0.6933 & 53.12\\
        8 & Q(0.75) & 128 & 2 & Q(0.75) & tanh & 0.3 & 128 & 0.6886 & 56.25\\
        9 & Q(0.75) & 128 & 4 & Q(0.75) & tanh & 0.3 & 64  & 0.6881 & 57.81\\
     \hline\hline
    \end{tabular}
    \end{center}
    \caption{Configuration specification of LSTM network with the different input op-code sequence lengths, embedding size, number of hidden layers, output dimension,activation function, drop-out ratio and batch size and the corresponding loss and accuracy(\%) on the validation data}
    \label{tbl:LSTM-configurations-used}
\end{table*}

In our experiments, Configuration-9 (C-9) produced the best accuracy. C-9 represents a setup of 4-layers of LSTM, each with 128 recurrent units. The input sequence was trimmed or padded to have an equal length across each instance. For C-9, this length was equal to the the $75^{th}$ percentile of total op-code sequence length distribution. Also, in configuration C-9 \textit{tanh} activation was used for the LSTM layers, a drop-out ratio of 0.3 was kept for each LSTM layer, and the batch-size was kept as 64. C-9 had the highest number of LSTM layers as used across the experimental design.
\par C-9's performance was closely followed by that of Configuration-6 (C-6). C-6 differed from C-9 in terms of the length of input sequence. It used $100^{th}$ percentile (or maximum of input-length) of op-code sequence length instead of $75^{th}$ percentile). The op-code sequences with less op-code than the $100^{th}$ percentile were padded to reach to this length. This required less number of LSTM layers (2 vs 4) and a lower size embedding (64 vs 128) and also a lower batch-size (32 vs 128). C-6 represented a much simpler configuration from an embedding and batch size perspective as compared to C-9, and the most complex configuration from the input-sequence-length perspective. 
It could also be inferred that the configurations with \textit{tanh} activation on an average performed better than \textit{sigmoid} activation.
    
\section{Discussion} \label{sec:discussion}
From the results obtained in section \ref{sec:results} it is evident that the highest accuracy is obtained by the LSTM configuration that had the maximum number of (hidden) layers. This observation indicates that there are many complex non-linear features hidden in the op-code sequences which require more LSTM layers for optimal feature representation. 
Another key observation is that the large size of the LSTM input (trimmed/ padded sequence length) and output dimension (nodes in the first flattened fully-connected layer), having number of nodes in the range of the length of the op-code sequence vector, is instrumental in achieving better performance.
This observation also indicates that the important op-code sub-sequence combinations which could help in detecting malware are spaced far apart in many cases. 
Besides these factors, the activation-function also plays a major role, and the networks with tanh activation in general demonstrated a better performance than the ones with sigmoid activation. 
We could also infer from the results that using a embedded inputs (by using embedding layer) instead of direct (raw) inputs is helpful for attaining better performance. An interesting observation is this regard is that a slightly higher embedding size may adequately compensate for a slightly lower input sequence length, but too high an embedding-size degrades the performance. This degradation may be due to the large (and hence complex) embeddings overfitting (on the noise in) the data.
Also as compared to the AE and DNN configuration for Malware detection as done by Sewak et. al. \cite{sewak-dnn}, the various Embedding-LSTM configurations used in our experiments did not produce a superior result with the same benchmark data. Hence, we cannot conclude that op-code sequence based DL (LSTM) for this dataset is superior than op-code frequency based DL (DNN) approach.

\section{Conclusion} \label{sec:conclusion}
Deep, recurrent language-models like the LSTM networks, have demonstrated exemplary performance in language-modeling. 
In this paper we demonstrated that for optimal results, these models as configured for spoken-language cannot optimally be used for machine-language (and malware detection) without major re-configurations. 
Using benchmark malware dataset, we experimented with different configurations of LSTM networks, and altered hyper-parameters like the embedding-size, number of hidden-layers, number of LSTM-units in a hidden layers, pruning/padding-length of the input-vector, activation-function, and batch-size.
We discovered that owing to the enhanced complexity of the malware/machine-language, the performance of an LSTM network that is configured for an Intrusion Detection System, is very sensitive towards the number-of-hidden-layers, input sequence-length and the choice of the activation-function. In the process we also identified the relative importance of all the key hyper-parameters of an LSTM for malware-detection and for designing an IDS.
\par 
We also questioned the assumption that as in the case of spoken-language-modeling, where the sequential DL networks like the LSTMs could by-far outperform their non-sequential counterparts like the MLP-DNNs, the same would be the case in the area of malware-language-modeling. 
We identified that for malware detection, the sequential DL architectures cannot out-perform their non-sequential counterparts on the same benchmark dataset. 
We believe that this work will be instrumental in sensitizing the research community over the precautions required while porting models and inferences from language-modeling domain for application in the cyber-security areas.


\begin{thebibliography}{8}
\bibitem{sahay}
Sanjay K Sahay, Ashu Sharma, Hemant Rathore, "Evolution of Malware and Its Detection Techniques", Information and Communication Technology for Sustainable Development, pp. 139--150, 2020.

\bibitem{1-Santos-2013}
I. Santos, F. Brezo, X. Ugarte-Pedrero, P. G. Bringas, `Opcode sequences as representation of executables for data-mining-based unknown malware detection', Information Sciences., vol 231, pp. 64\textemdash82, 2013.

\bibitem{2-Ashu-RF}
Ashu Sharma, Sanjay Kumar Sahay, Abhishek Kumar, `Improving the Detection Accuracy of Unknown Malware by Partitioning the Executables in Groups', Advanced Computing and Communication Technologies,Springer Singapore, pp. 421\textemdash431, 2016.

\bibitem{3-Malicia}
Malicia project, http://malicia-project.com/, accessed in 2014.

\bibitem{sewak-dnn}
Mohit Sewak, Sanjay K. Sahay and Hemant Rathore, An investigation of a deep learning based malware detection system, CoRR, arXiv, volume abs/1809.05888, 2018.

\bibitem{sewak-comparison}
Mohit Sewak, Sanjay K. Sahay and Hemant Rathore, Comparison of Deep Learning and the Classical Machine Learning Algorithm for the Malware Detection, CoRR, volume abs/1809.05889, 2018.

\bibitem{LSTM}
Sepp Hochreiter, Jürgen Schmidhuber, "Long Short-Term Memory". Neural Computation. 9 (8), pp. 1735--1780, 1997.

\bibitem{20-DeepSign-2015}
O. E. David, N. S. Netanyahu, `DeepSign: Deep learning for automatic malware signature generation and classification',2015 International Joint Conference on Neural Networks (IJCNN), Killarney, pp. 1\textemdash8, 2015.

\bibitem{21-Saxe-2015}
J. Saxe, K. Berlin, `Deep neural network based malware detection using two dimensional binary program features', 2015 10th International Conference on Malicious and Unwanted Software (MALWARE), Fajardo, pp. 11\textemdash20, 2015.

\bibitem{hemant-bda}
Hemant Rathore, Swati Agarwal, Sanjay K Sahay, Mohit Sewak, "Malware Detection Using Machine Learning and Deep Learning", International Conference on Big Data Analytics (BDA), pp. 402-411, 2018.

\bibitem{19-Pascanu-2015}
R. Pascanu et. al., `Malware classification with recurrent networks', In 2015 IEEE International Conference on Acoustics, Speech and Signal Processing (ICASSP), pp. 1916--1920. 2015.

\bibitem{opcode-lstm-2layer}
Renjie Lu, "Malware Detection with {LSTM} using Opcode Language", CoRR, volume abs/1906.04593, 2019

\bibitem{microsoft-lstm}
B. Athiwaratkun and J. W. Stokes, "Malware classification with LSTM and GRU language models and a character-level CNN", 2017 IEEE International Conference on Acoustics, Speech and Signal Processing (ICASSP), pp. 2482--2486, 2017.

\bibitem{22-Kolosnjaji-2016}
Kolosnjaji Bojan et. al., `Deep Learning for Classification of Malware System Call Sequences', AI 2016: Advances in Artificial Intelligence: 29th Australasian Joint Conference, 2016, Proceedings", Pg. 137\textemdash149, 2016.

\bibitem{18-Xin-2016}
Xin Wang, Siu Ming Yiu, `A multi-task learning model for malware classification with useful file access pattern from API call sequence', CoRR, arXiv:1610.05945, 2016.

\bibitem{LSTM-api-call}
S. Maniath et. al., "Deep learning LSTM based ransomware detection," 2017 Recent Developments in Control, Automation \& Power Engineering (RDCAPE), pp. 442--446, 2017.

\bibitem{LSTM-system-call}
Xi Xiao et. al., "Android malware detection based on system call sequences and LSTM", Multimedia Tools and Applications, pp. 3979--3999, 2019.

\bibitem{LSTM-PE-header}
Edward Raff et. al., "Learning the PE Header, Malware Detection with Minimal Domain Knowledge", In Proceedings of the 10th ACM Workshop on Artificial Intelligence and Security (AISec ’17), 2017.

\bibitem{hemant-cluster}
Hemant Rathore, Sanjay K Sahay, Palash Chaturvedi, Mohit Sewak, "Android Malicious Application Classification Using Clustering", International Conference on Intelligent Systems Design and Applications (ISDA), pp. 659--667, 2018.

\bibitem{LSTM-iot}
Hamed Haddad Pajouh et. al., "A deep Recurrent Neural Network based approach for Internet of Things malware threat hunting", Future Generation Computer Systems, vol. 85, pp. 88--96, 2018.

\bibitem{LSTM-obfuscation}.
Alessandro Bacci et.al, "Detection of Obfuscation Techniques in Android Applications", In Proceedings of the 13th International Conference on Availability, Reliability and Security (ARES 2018), 2018.

\bibitem{LSTM-HDN}
Jinpei Yan et. al., "LSTM-Based Hierarchical Denoising Network for Android Malware Detection", Security and Communication Networks, 2018.

\bibitem{lstm-original-paper}
Sepp Hochreiter; Jürgen Schmidhuber, "Long short-term memory", Neural Computation, 9 (8), pp. 1735--1780, 1997.

\bibitem{lstm-forget-gate}
Klaus Greff et. al., "LSTM: A Search Space Odyssey". IEEE Transactions on Neural Networks and Learning Systems, 28 (10), pp. 2222-–2232, 2015.

\bibitem{lstm-forget-forward-pass}
Felix A. Gers, Jürgen Schmidhuber, Fred Cummins, "Learning to Forget: Continual Prediction with LSTM", Neural Computation, 12 (10), pp. 2451--2471, 2000.

\bibitem{lstm-hadamard-product}
Chandler Davis, "The norm of the Schur product operation". Numerische Mathematik, 4 (1), pp. 343--44, 1962.

\bibitem{sewak-dl-overview}
Mohit Sewak, Sanjay K. Sahay and Hemant Rathore, "An Overview of Deep Learning Architecture of Deep Neural Networks and Autoencoders", Journal of Computational and Theoretical Nanoscience, 2018.

\bibitem{lstm-peephole-activation}
F. A. Gers, J. Schmidhuber "LSTM Recurrent Networks Learn Simple Context Free and Context Sensitive Languages", IEEE Transactions on Neural Networks. 12 (6), pp. 1333--1340, 2001. doi:10.1109/72.963769ain






\end{thebibliography}

\end{document}